\documentclass[oribibl]{llncs}
\usepackage{graphics}
\newcommand{\Prob}[1]{\ensuremath{\mathrm{Pr}\left( #1 \right)}}
\newcommand{\CSSR}{\texttt{CSSR}}
\begin{document}

\title{Discovering Functional Communities in Dynamical Networks}
\author{Cosma Rohilla Shalizi\inst{1} \and Marcelo F. Camperi\inst{2} \and Kristina Lisa Klinkner\inst{1}}
\institute{Statistics Department, Carnegie Mellon University, Pittsburgh, PA 15213 USA \email{cshalizi@cmu.edu,klinkner@cmu.edu}
\and Physics Department, University of San Francisco, San Francisco, CA 94118 USA
\email{camperi@usfca.edu}}
\maketitle

\begin{abstract}
Many networks are important because they are substrates for dynamical systems,
and their pattern of functional connectivity can itself be dynamic --- they can
functionally reorganize, even if their underlying anatomical structure remains
fixed.  However, the recent rapid progress in discovering the community
structure of networks has overwhelmingly focused on that constant anatomical
connectivity.  In this paper, we lay out the problem of discovering {\em
  functional communities}, and describe an approach to doing so.  This method
combines recent work on measuring information sharing across stochastic
networks with an existing and successful community-discovery algorithm for
weighted networks.  We illustrate it with an application to a large biophysical
model of the transition from beta to gamma rhythms in the hippocampus.
\end{abstract}

\section{Introduction}

The {\em community discovery} problem for networks is that of splitting a
graph, representing a group of interacting processes or entities, into
sub-graphs (communities) which are somehow modular, so that the nodes belonging
to a given sub-graph interact with the other members more strongly than they do
with the rest of the network.  As the word ``community'' indicates, the problem
has its roots in the study of social structure and cohesion
\cite{Simmel-web,Scott-social-network-analysis,Moody-White-social-cohesion},
but is related to both general issues of clustering in statistical data mining
\cite{Hand-Mannila-Smyth} and to the systems-analysis problem of decomposing
large systems into weakly-coupled sub-systems
\cite{Simon-architecture,Alexander-notes,Siljak-decentralized-control}.

The work of Newman and Girvan \cite{MEJN-Girvan-community-structure} has
inspired a great deal of research on statistical-mechanical approaches to
community detection in complex networks.  (For a recent partial review, see
\cite{MEJN-community-structure-from-eigenvectors}.)  To date, however, this
tradition has implicitly assumed that the network is defined by persistent, if
not static, connections between nodes, whether through concrete physical
channels (e.g., electrical power grids, nerve fibers in the brain), or through
enduring, settled patterns of interaction (e.g., friendship and collaboration
networks).  However, networks can also be defined through {\em coordinated
  behavior}, and the associated sharing of dynamical information; neuroscience
distinguishes these as, respectively, ``anatomical'' and ``functional''
connectivity
\cite{Sporns-Tononi-Edelman-theoretical-neuroanatomy,Friston-beyond-phrenology}.
The two sorts of connectivity do not map neatly onto each other, and it would
be odd if functional modules always lined up with anatomical ones.  Indeed, the
same system could have many different sets of functional communities in
different dynamical regimes.  For an extreme case, consider globally-coupled
map lattices \cite{Kaneko-globally-coupled-maps}, which are important
statistical-mechanical models of physical and biological pattern formation
\cite{Chazottes-Fernandez-on-CML}.  In these systems, the ``anatomical''
network is fully connected, so there is only a single (trivial) community.
Nonetheless, in some dynamical regimes they spontaneously develop many
functional communities, i.e., groups of nodes which are internally coherent but
with low inter-group coordination
\cite{Delgado-Sole-turbulence-and-finite-automata}.\footnote{We are preparing a
  separate paper on functional communities in coupled map lattices.}

\vspace{-0.5mm} Coupled map lattices are mathematical models, but the
distinction between anatomical and functional communities is not merely a
conceptual possibility.  Observation of neuronal networks {\em in vivo} show
that it is fairly common for, e.g., central pattern generators to change their
functional organization considerably, depending on which pattern they are
generating, while maintaining a constant anatomy \cite{Crustacean-STG-as-book}.
Similarly, neuropsychological evidence has long suggested that there is no
one-to-one mapping between higher cognitive functions and specialized cortical
modules, but rather that the latter participate in multiple functions and vice
versa, re-organizing depending on the task situation
\cite{Luria-working-brain}.  Details of this picture, of specialized anatomical
regions supporting multiple patterns of functional connectivity, have more
recently been filled in by brain imaging studies
\cite{Friston-beyond-phrenology}.  Similar principles are thought to govern the
immune response, cellular signaling, and other forms of biological information
processing \cite{Segel-Cohen-immune-system}.  Thus, in analyzing these
biological networks, it would be highly desirable to have a way of detecting
functional communities, rather than just anatomical ones.  Similarly, while
much of the work on social network organization concerns itself with the
persistent ties which are analogous to anatomy, it seems very likely
\cite{Chisholm-coordination-without-hierarchy,Luhmann-social-systems} that
these communities cut in complicated ways across the functional ones defined by
behavioral coordination \cite{Lindblom-intelligence,Young-strategy-structure}
or information flow \cite{Eckmann-Moses-Sergi-dialog}.  This is perhaps
particularly true of modern societies, which are thought, on several grounds
\cite{Coser-modernity,Gellner-plough-sword-book,Luhmann-social-systems} to be
more flexibly organized than traditional ones.

In this paper, we propose a two-part method to discover functional communities
in network dynamical systems.  Section \ref{sec:info-coh} describes the first
part, which is to calculate, across the whole of the network, an appropriate
measure of behavioral coordination or information sharing; we argue that
informational coherence, introduced in our prior work \cite{Info-Coh-for-NIPS},
provides such a measure.  Section \ref{sec:community-discovery} describes the
other half of our method, using our measure of coordination in place of a
traditional adjacency matrix in a suitable community-discovery algorithm.  Here
we employ the Potts model procedure proposed by Reichardt and Bornholdt
\cite{Reichardt-Bornholdt-community-structures,Reichardt-Bornholdt-stat-mech-of-community-detection}.
Section \ref{sec:method-summary} summarizes the method and clarifies the
meaning of the functional communities it finds.  Sections \ref{sec:model} and
\ref{sec:results-on-model} apply our method to a detailed biophysical model of
collective oscillations in the hippocampus
\cite{New-roles-for-the-gamma-rhythm}, where it allows us to detect the
functional re-organization accompanying the transition from gamma to beta
rhythms.  Finally, Sect. \ref{sec:discussion} discusses the limitations of our
method and its relations to other approaches (Sect.\
\ref{sec:limitations-and-relations}) and some issues for future work (Sect.\
\ref{sec:future-work}).

\section{Discovering Behavioral Communities}

There are two parts to our method for finding functional communities.  We first
calculate a measure of the behavioral coordination between all pairs of nodes
in the network: here, the informational coherence introduced in
\cite{Info-Coh-for-NIPS}.  We then feed the resulting matrix into a suitable
community-discovery algorithm, in place of the usual representation of a
network by its adjacency matrix.  Here, we have used the Reichardt-Bornholdt
algorithm \cite{Reichardt-Bornholdt-community-structures}, owing to its
Hamiltonian form, its ability to handle weighted networks, and its close
connection to modularity.

\subsection{Informational Coherence}
\label{sec:info-coh}

We introduced {\em informational coherence} in \cite{Info-Coh-for-NIPS} to
measure the degree to which the behavior of two systems is coordinated, i.e.,
how much dynamically-relevant information they share.  Because of its
centrality to our method, we briefly recapitulate the argument of that paper.

The starting point is the strong notion of ``state'' employed in physics and
dynamical systems theory: the state of the system is a variable which
determines the distribution of all present and future observables.  In
inferential terms, the state is a minimal sufficient statistic for predicting
future observations \cite{CMPPSS}, and can be formally constructed as
measure-valued process giving the distribution of future events conditional on
the history of the process.  As a consequence, the state always evolves
according to a homogeneous Markov process \cite{Knight-predictive-view,CMPPSS}.

In a dynamical network, each node $i$ has an associated time-series of
observations $X_i(t)$.  This is in turn generated by a Markovian state process,
$S_i(t)$, which forms its optimal nonlinear predictor.  For any two nodes $i$
and $j$, the informational coherence is
\begin{equation}
\label{eqn:ic-defn} IC_{ij} \equiv \frac{I[S_i;S_j]}{\min{H[S_i],H[S_j]}}
\end{equation}
where $I[S_i;S_j]$ is the mutual information shared by $S_i$ and $S_j$, and
$H[S_i]$ is the self-information (Shannon entropy) of $S_i$.  Since $I[S_i;S_j]
\leq \min{H[S_i],H[S_j]}$, this is a symmetric quantity, normalized to lie
between 0 and 1 inclusive.  The construction of the predictive states ensures
that $S_i(t)$ encapsulates all information in the past of $X_i(t)$ which is
relevant to its future, so a positive value for $I[S_i;S_j]$ means that
$S_j(t)$ contains information about the future of $X_i(t)$.  That is, a
positive value of $I[S_i;S_j]$ is equivalent to the sharing of {\em dynamically
  relevant} information between the nodes, manifesting itself as coordinated
behavior on the part of nodes $i$ and $j$.

Clearly, a crucial step in calculating informational coherence is going from
the observational time series $X_i(t)$ to the predictive state series $S_i(t)$.
In certain cases with completely specified probability models, this can be done
analytically \cite{CMPPSS,predictive-representations-of-state}.  In general,
however, we are forced to reconstruct the appropriate state-space structure
from the time series itself.  State reconstruction for deterministic systems is
based on the Takens embedding theorem, and is now routine
\cite{Kantz-Schreiber}.  However, biological and social systems are hardly ever
deterministic at experimentally-accessible levels of resolution, so we need a
stochastic state reconstruction algorithm.  Several exist; we use the
\CSSR\  algorithm introduced in \cite{CSSR-for-UAI}, since, so far as we
know, it is currently the only stochastic state reconstruction algorithm which
has been proved statistically consistent (for conditionally stationary discrete
sequences).  We briefly describe \CSSR\  in Appendix \ref{app:CSSR}.

Informational coherence is not, of course, the only possible way of measuring
behavioral coordination, or functional connectivity.  However, it has a number
of advantages over rival measures \cite{Info-Coh-for-NIPS}.  Unlike measures of
strict synchronization, which insist on units doing exactly the same thing at
exactly the same time, it accommodates phase lags, phase locking, chaotic
synchronization, etc., in a straightforward and uniform manner.  Unlike
cross-covariance, or the related spectral coherence, it easily handles
nonlinear dependencies, and does not require the choice of a particular lag (or
frequency, for spectral coherence), because the predictive states summarize the
entire relevant portion of the history.  Generalized synchrony measures
\cite{Quian-Quiroga-Grassberger-synch} can handle nonlinear relationships among
states, but inappropriately assume determinism.  Finally, mutual information
among the observables, $I[X_i;X_j]$, can handle nonlinear, stochastic
dependencies, but suffers, especially in neural systems, because what we really
want to detect are coordinated {\em patterns} of behavior, rather than
coordinated instantaneous actions.  Because each predictive state corresponds
to a unique statistical pattern of behavior, mutual information among these
states is the most natural way to capture functional connectivity.

\subsection{The Reichardt-Bornholdt Community Discovery Algorithm}
\label{sec:community-discovery}

The Reichardt-Bornholdt \cite{Reichardt-Bornholdt-community-structures,%
  Reichardt-Bornholdt-stat-mech-of-community-detection} community discovery
algorithm finds groups of nodes that are densely coupled to one another, but
only weakly coupled to the rest of the network, by establishing a (fictitious)
spin system on the network, with a Hamiltonian with precisely the desired
properties, and then minimizing the Hamiltonian through simulated annealing.
More concretely, every node $i$ is assigned a ``spin'' $\sigma_i$, which is a
discrete variable taking an integer value from $1$ to a user-defined $q$.  A
``community'' or ``module'' will consist of all the nodes with a common spin
value.  The spin Hamiltonian combines a ferromagnetic term, which favors linked
nodes taking the same spin (i.e., being in the same community), and an
anti-ferromagnetic term, which favors non-linked nodes taking different spins
(i.e., being in different community).  Both interactions are of the Potts model
type, i.e., they are invariant under permutations of the integers labeling the
clusters.  After some algebraic manipulation
\cite{Reichardt-Bornholdt-stat-mech-of-community-detection}, one arrives at the
Hamiltonian
\begin{equation}
\mathcal{H}(\sigma) = -\sum_{i \neq j}{(A_{ij} - \gamma p_{ij}) \delta(\sigma_i,\sigma_j)}
\label{eqn:r-b-hamiltonian}
\end{equation}
where $A_{ij}$ is the adjacency matrix, $\delta(\cdot,\cdot)$ is the Kronecker
delta function, $p_{ij}$ is a matrix of non-negative constants giving the
relative weights of different possible links, and $\gamma$ gives the relative
contribution of link absence to link presence.  The choice of $p_{ij}$ is
actually fairly unconstrained, but previous experience with community discovery
suggests that very good results are obtained by optimizing the Newman
modularity $Q$ \cite{MEJN-mixing-patterns}
\begin{equation}
Q(\sigma) = \frac{1}{2M}\sum_{i, j}{\left(A_{ij} - \frac{k_i k_j}{2M}\right) \delta(\sigma_i,\sigma_j)}
\label{eqn:mejn-modularity}
\end{equation}
where $k_i$ is the degree of node $i$, and $2M = \sum_{i}{k_i}$ the total number
of links.  Essentially, Newman's $Q$ counts the number of edges within
communities, minus the number which would be expected in a randomized graph
where each node preserved its actual degree
\cite{MEJN-community-structure-from-eigenvectors}, and $\sigma_i$ were IID
uniform.  Setting $p_{ij} = k_i k_j/2M$ and $\gamma = 1$, we see that
$\mathcal{H}(\sigma)$ and $-Q(\sigma)$ differ only by a term (the diagonal part
of the sum for $Q$) which does not depend on the assignment of nodes to
communities.  Thus, minimizing $\mathcal{H}(\sigma)$ is the same as maximizing
the modularity.  Varying $\gamma$, in this scheme, effectively controls the
trade-off between having many small communities and a few large ones
\cite{Reichardt-Bornholdt-stat-mech-of-community-detection}, and makes it
possible to discover a hierarchical community structure, which will be the
subject of future work.

While this procedure was originally developed for the case where $A_{ij}$ is a
$0$-$1$ adjacency matrix, it also works perfectly well when links take on
(positive) real-valued strengths.  In particular, using $A_{ij} = IC_{ij}$, we
can still maximize the modularity, taking the ``degree'' of node $i$ to be $k_i
= \sum_{j}{IC_{ij}}$
\cite{Reichardt-Bornholdt-stat-mech-of-community-detection}.  The
interpretation of the modularity is now the difference between the strength of
intra-community links, and a randomized model where each node shares its link
strength indifferently with members of its own and other communities.

\subsection{Summary of the Method}
\label{sec:method-summary} 

Let us briefly summarize the method for discovering functional communities.  We
begin with a network, consisting of $N$ nodes.  For each node, we have a
discrete-value, discrete-time (``symbolic'') time series,
$\left\{x_i(t)\right\}$, recorded simultaneously over all nodes.  The
\CSSR\ algorithm is applied to each node's series separately, producing a set
of predictive states for that node, and a time series of those states, $\left\{
s_i(t)\right\}$.  We then calculate the complete set of pairwise informational
coherence values, $\left\{IC_{ij}\right\}$, using Eq.\ \ref{eqn:ic-defn}.  This
matrix is fed into the Reichardt-Bornholdt procedure, with $A_{ij} = IC_{ij}$,
which finds an assignment of spins to nodes, $\left\{\sigma_i\right\}$,
minimizing the Hamiltonian in Eq.\ \ref{eqn:r-b-hamiltonian}.  The functional
communities of the dynamical network consist of groups of nodes with common
spin values.  Within each community, the average pairwise coherence of the
nodes is strictly greater than would be expected from a randomizing null model
(as described in the previous paragraph).  Furthermore, between any two
communities, the average pairwise coherence of their nodes is strictly less
than expected from randomization
\cite{Reichardt-Bornholdt-stat-mech-of-community-detection}.

\section{Test on a Model System of Known Structure: Collective Oscillations in the Hippocampus}
\label{sec:model}

We use simulated data as a test case, to validate the general idea of our
method, because it allows us to work with a substantial network where we
nonetheless have a strong idea of what appropriate results should be.  Because
of our ultimate concern with the functional re-organization of the brain, we
employed a large, biophysically-detailed neuronal network model, with over 1000
simulated neurons.

The model, taken from \cite{New-roles-for-the-gamma-rhythm}, was originally
designed to study episodes of gamma (30--80Hz) and beta (12--30Hz) oscillations
in the mammalian nervous system, which often occur successively with a
spontaneous transition between them. More concretely, the rhythms studied were
those displayed by {\it in vitro} hippocampal (CA1) slice preparations and by
{\it in vivo} neocortical EEGs.

\begin{figure}
\begin{center}$a$\resizebox{0.9\columnwidth}{!}{\includegraphics{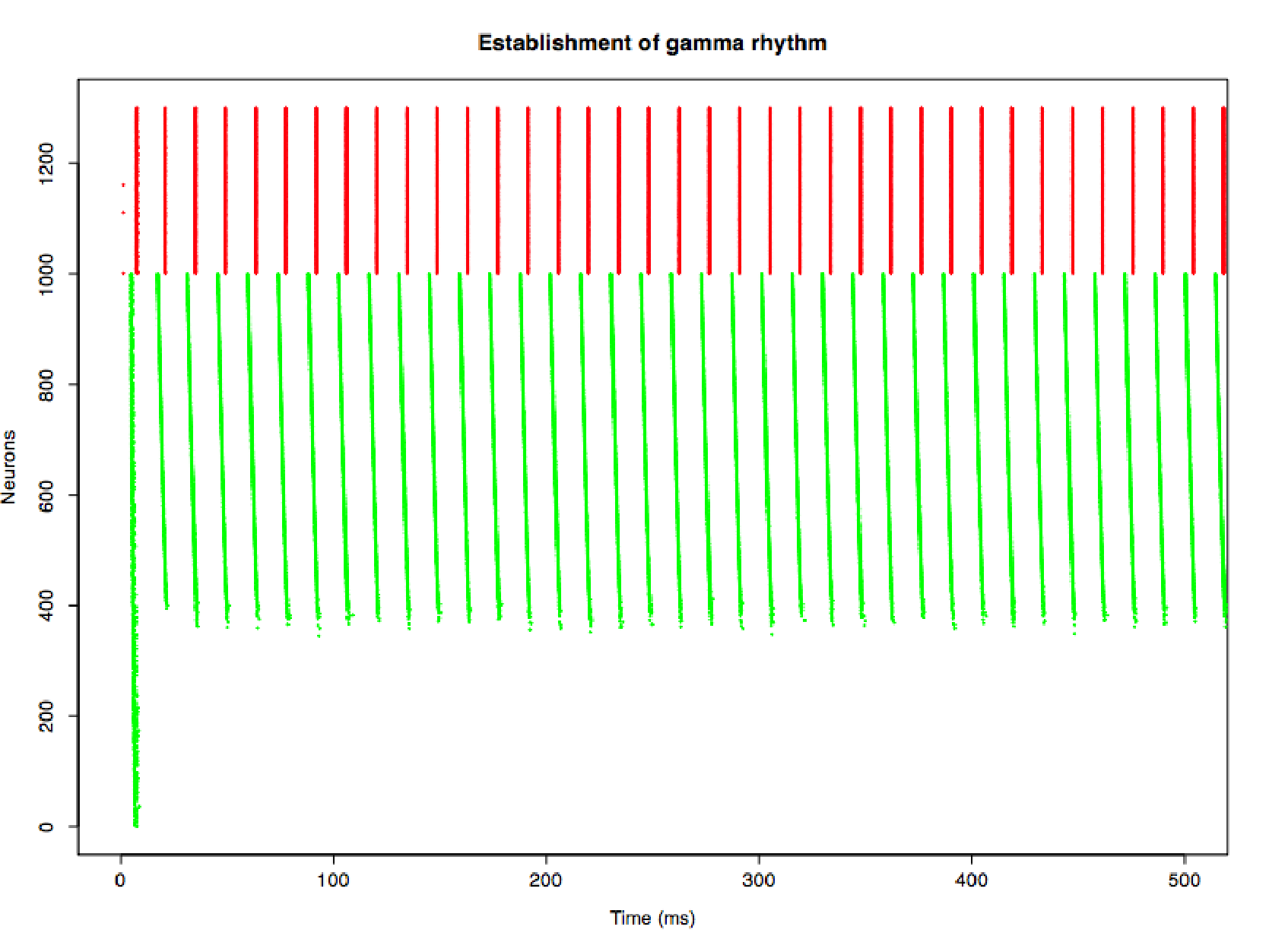}} $b$\resizebox{0.9\columnwidth}{!}{\includegraphics{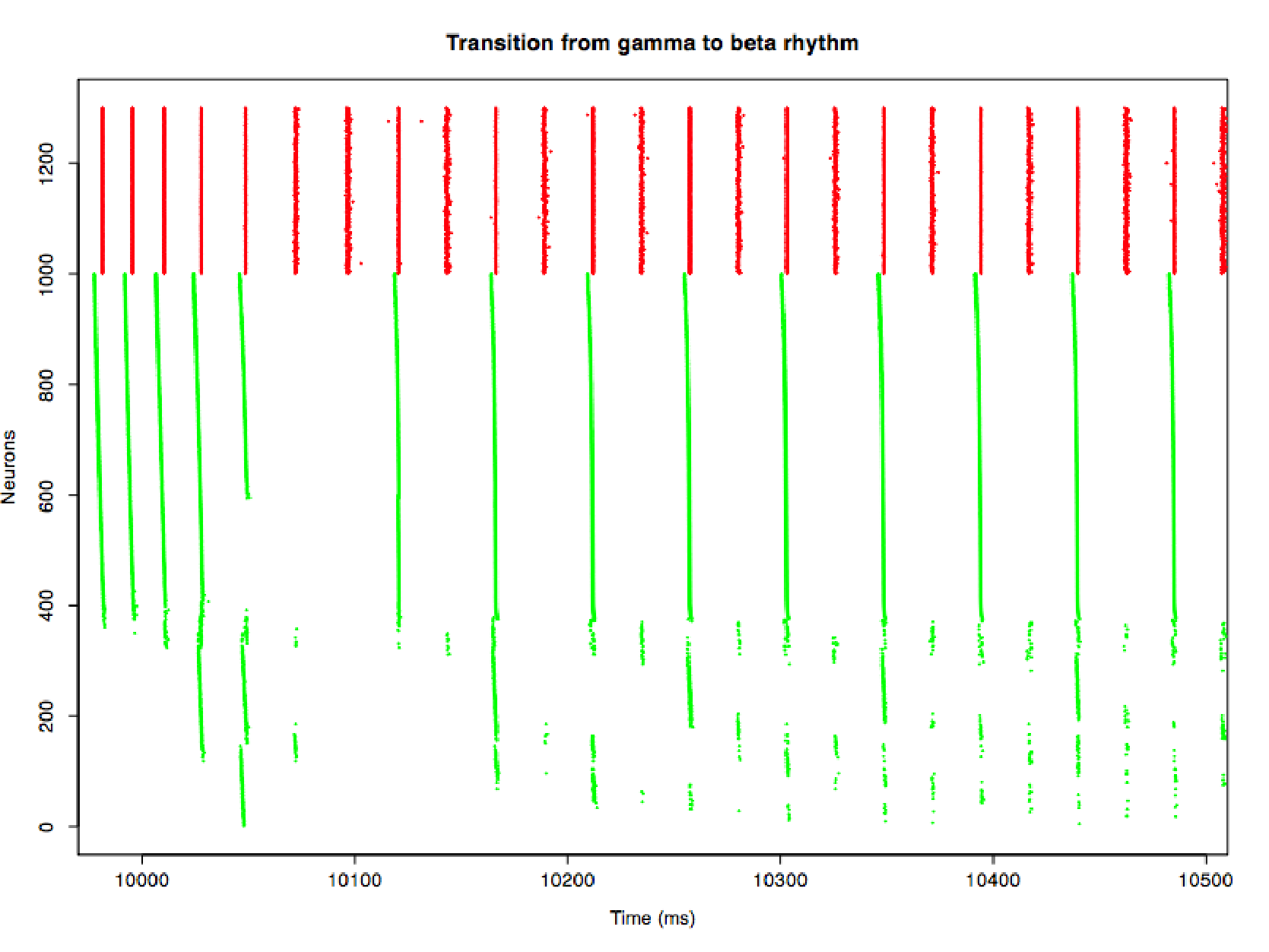}}\end{center}
\caption{Rastergrams of neuronal spike-times in the network.  Excitatory,
  pyramidal neurons (numbers 1 to 1000) are green, inhibitory interneurons
  (numbers 1001 to 1300) are red.  During the first 10 seconds ($a$), the
  current connections among the pyramidal cells are suppressed and a gamma
  rhythm emerges (left).  At $t=10\mathrm{s}$, those connections become active,
  leading to a beta rhythm ($b$, right).}
\label{fig:rastergram}
\end{figure}

\begin{figure}[t]
\begin{center}$a$\resizebox{0.5\columnwidth}{!}{\includegraphics{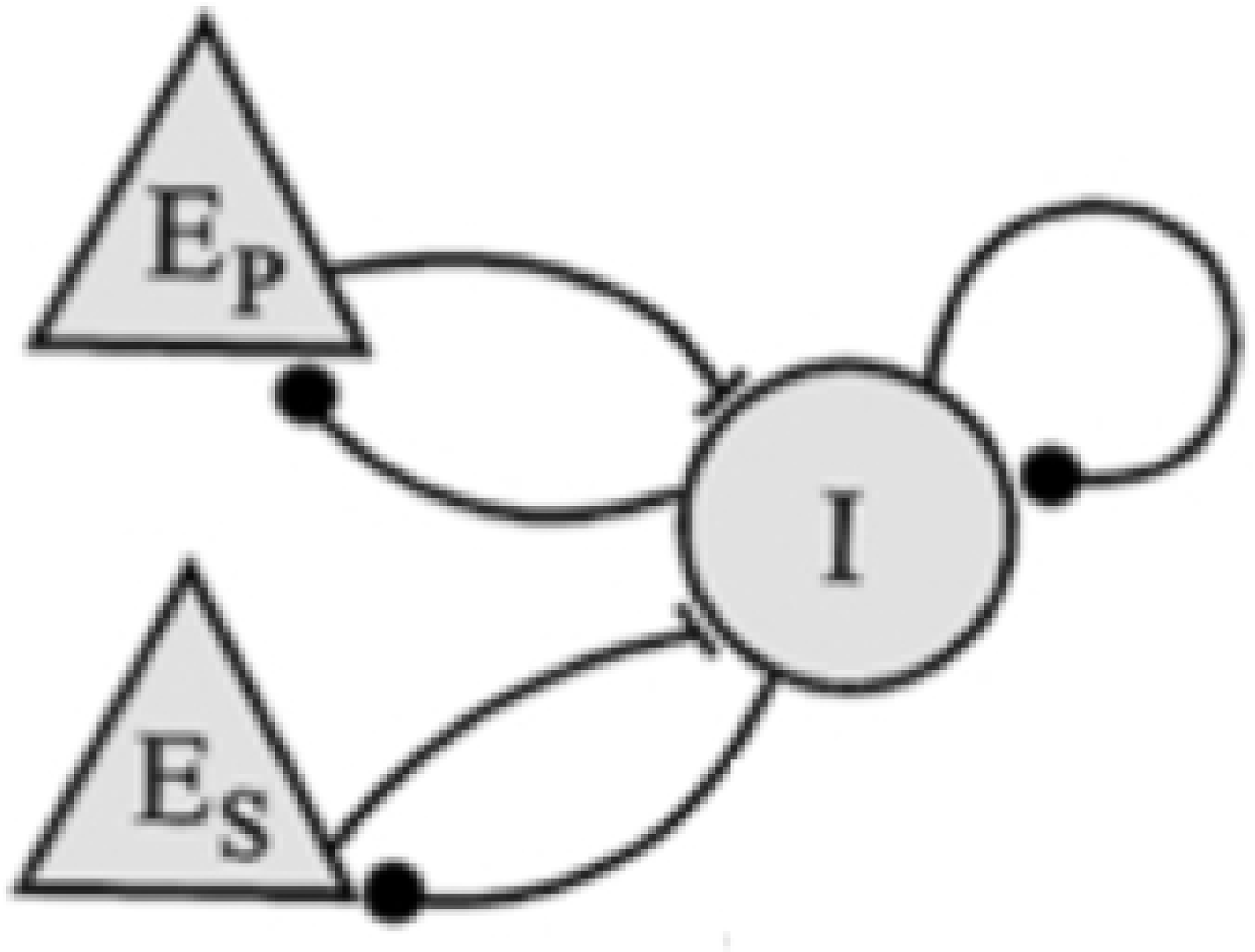}} $b$\resizebox{0.4\columnwidth}{!}{\includegraphics{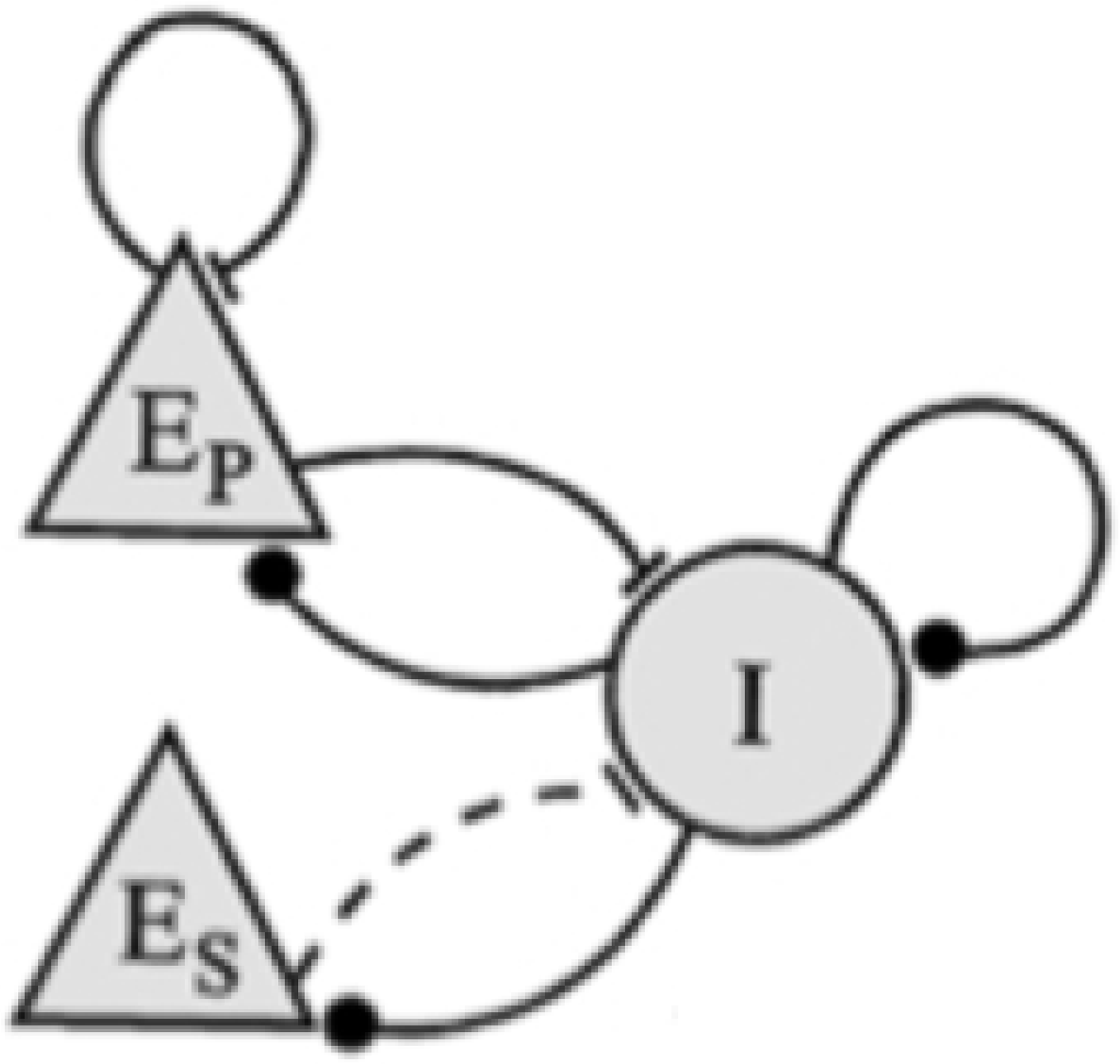}}\end{center}
\caption{Schematic depiction of the anatomical network.  Here nodes represent
  populations of cells: excitatory pyramidal neurons (triangles labeled
  $\mathrm{E}$) or inhibitory interneurons (large circle labeled $\mathrm{I}$).
  Excitatory connections terminate in bars, inhibitory connections in filled
  circles.  During the gamma rhythm ($a$), the pyramidal neurons are coupled to
  each other only indirectly, via the interneurons, and dynamical effects
  separate the pyramidal population into participating ($\mathrm{E_P}$) and
  suppressed ($\mathrm{E_S}$) sub-populations.  During the beta rhythm ($b$),
  direct connections among the $\mathrm{E_P}$ neurons, built up, but not
  activated, by Hebbian learning under the gamma rhythm are turned on, and the
  connection from the $\mathrm{E_S}$ neurons to the interneurons are weakened
  by the same Hebbian process (dashed line).}
\label{fig:anatomical-connections}
\end{figure}

The model contains two neuron populations: excitatory (AMPA) pyramidal neurons
and inhibitory (GABA$_{\mbox{\footnotesize{A}}}$) interneurons, defined by
conductance-based Hodgkin-Huxley-style equations.  Simulations were carried out
in a network of 1000 pyramidal cells and 300 interneurons.  Each cell was
modeled as a one-compartment neuron with all-to-all coupling, endowed with the
basic sodium and potassium spiking currents, an external applied current, and
some Gaussian input noise.  The anatomical, synaptic connections were organized into blocks, as shown in Fig.\ \ref{fig:anatomical-connections}.

The first 10 seconds of the simulation correspond to the gamma rhythm, in which
only a group of neurons is made to spike via a linearly increasing applied
current.  The beta rhythm (subsequent 10 seconds) is obtained by activating
pyramidal-pyramidal recurrent connections (potentiated by Hebbian preprocessing
as a result of synchrony during the gamma rhythm) and a slow outward
after-hyper-polarization (AHP) current (the M-current), suppressed during gamma
due to the metabotropic activation used in the generation of the rhythm.
During the beta rhythm, pyramidal cells, silent during gamma rhythm, fire on a
subset of interneurons cycles (Fig.\ \ref{fig:rastergram}).

\begin{figure}
\begin{center}$a$\resizebox{0.8\columnwidth}{!}{\includegraphics{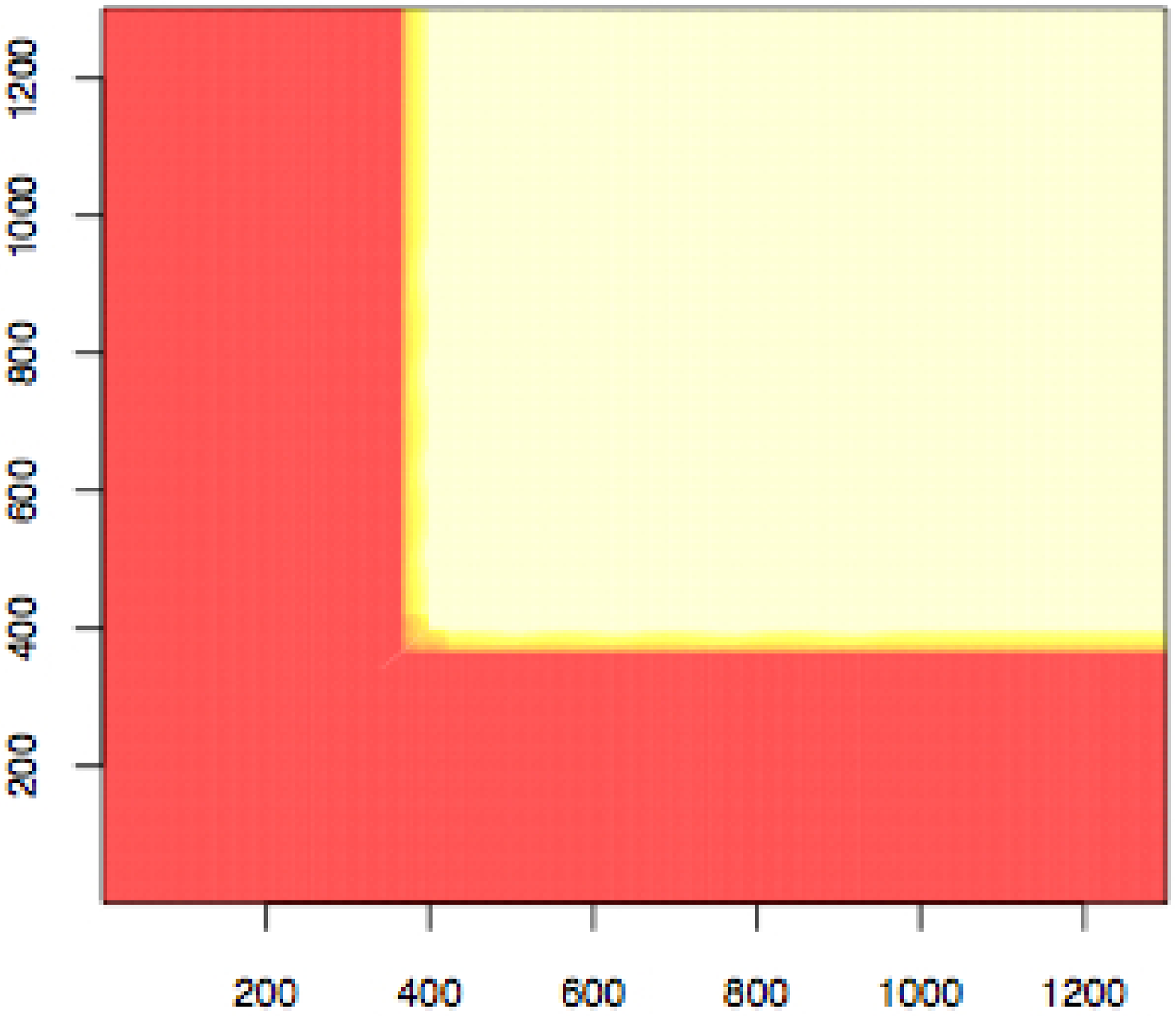}} $b$\resizebox{0.8\columnwidth}{!}{\includegraphics{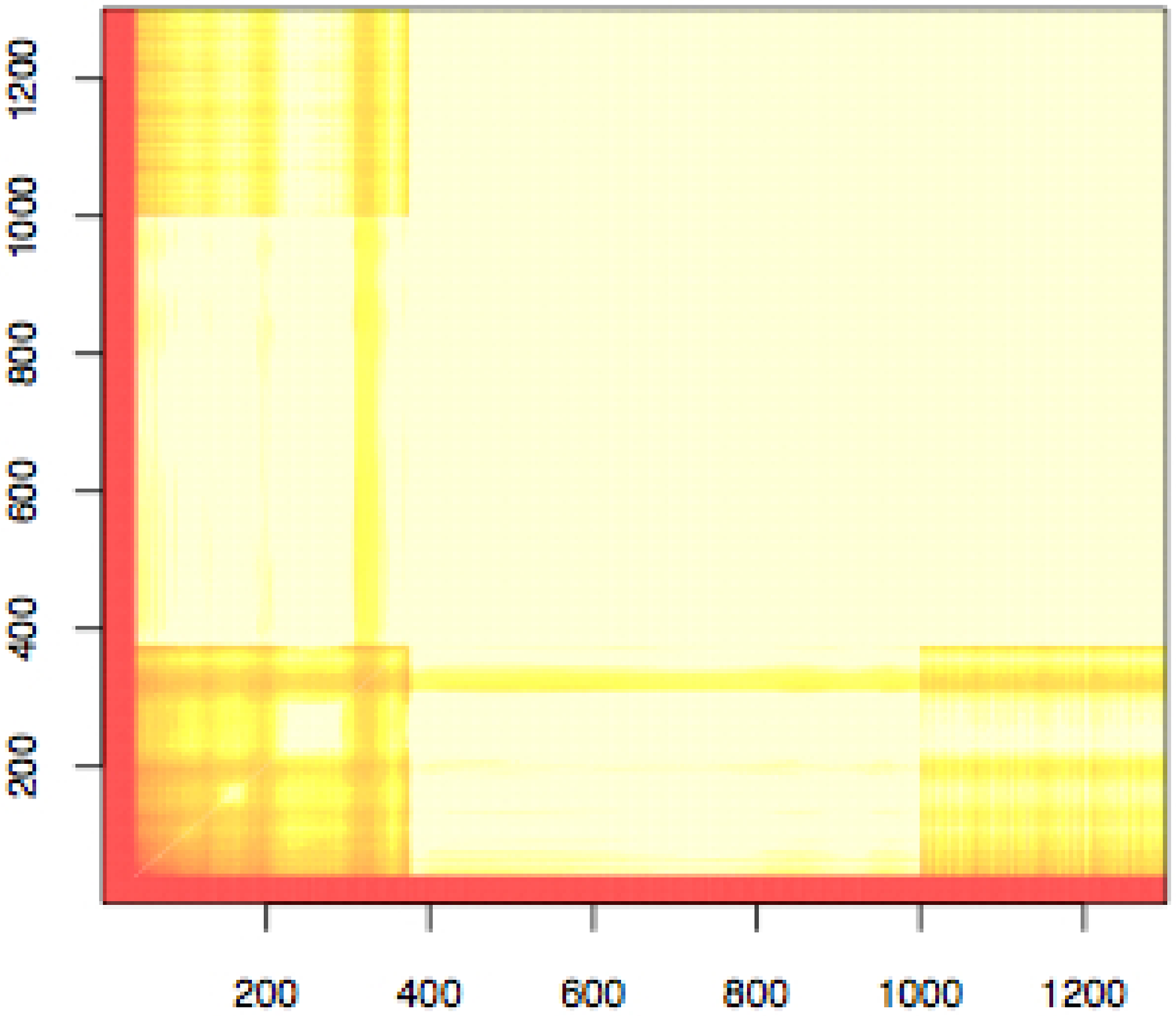}}
\end{center}
\caption{Heat-maps of coordination across neurons in the network, measured by
  informational coherence.  Colors run from red (no coordination) through
  yellow to pale cream (maximum).}
\label{fig:heatmap}
\end{figure}

\section{Results on the Model}
\label{sec:results-on-model}

A simple heat-map display of the informational coherence
(Fig.\ \ref{fig:heatmap}) shows little structure among the active neurons in
either regime.  However, visual inspection of the rastergrams
(Fig. \ref{fig:rastergram}) leads us to suspect the presence of two very large
functional communities: one, centered on the inhibitory interneurons and the
excitatory pyramidal neurons most tightly coupled to them, and another of the
more peripheral excitatory neurons.  During the switch from the gamma to the
beta rhythm, we expect these groups to re-organize.

\begin{figure}
\begin{center}$a$\resizebox{0.62\columnwidth}{!}{\includegraphics{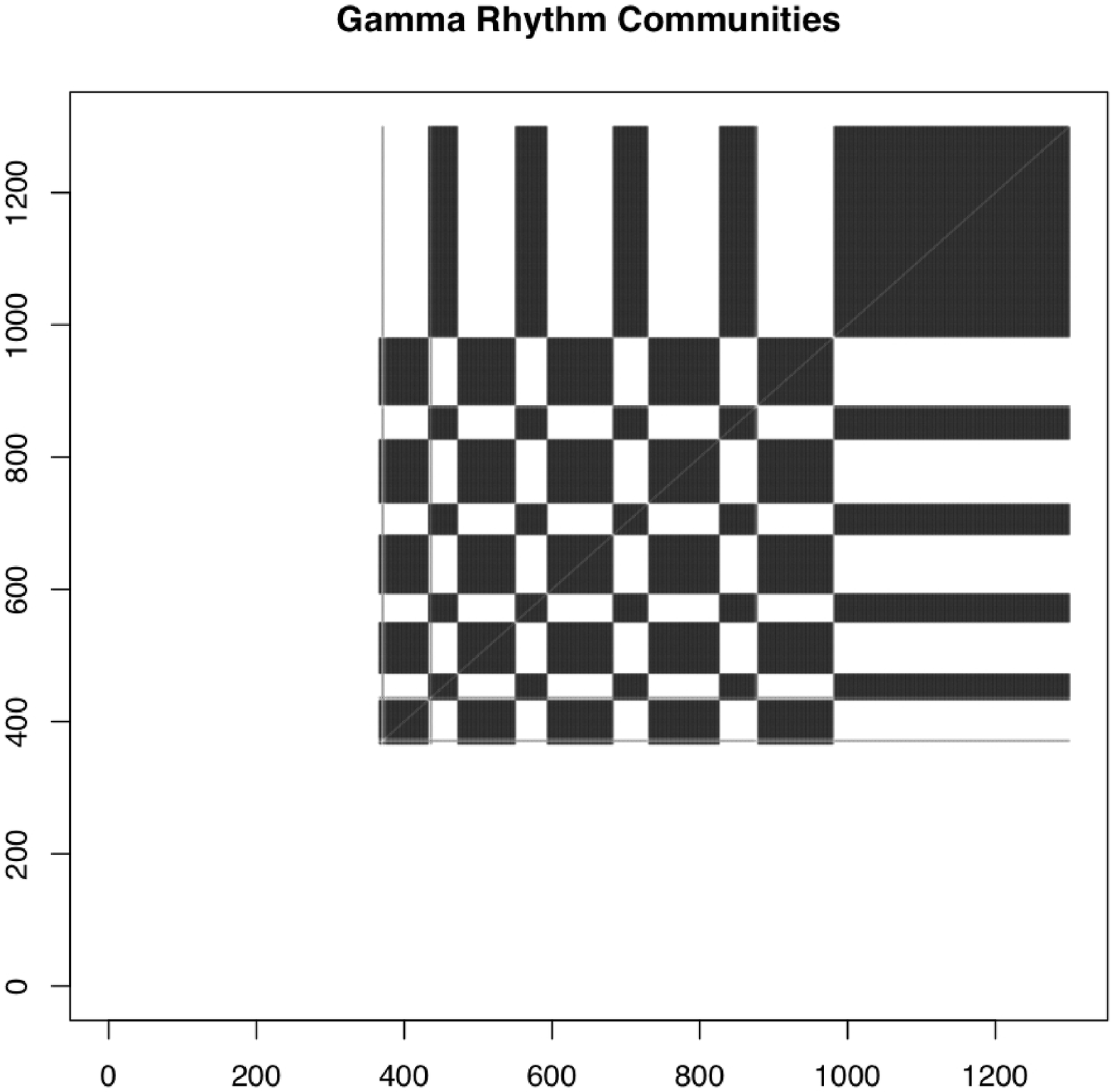}} $b$\resizebox{0.62\columnwidth}{!}{\includegraphics{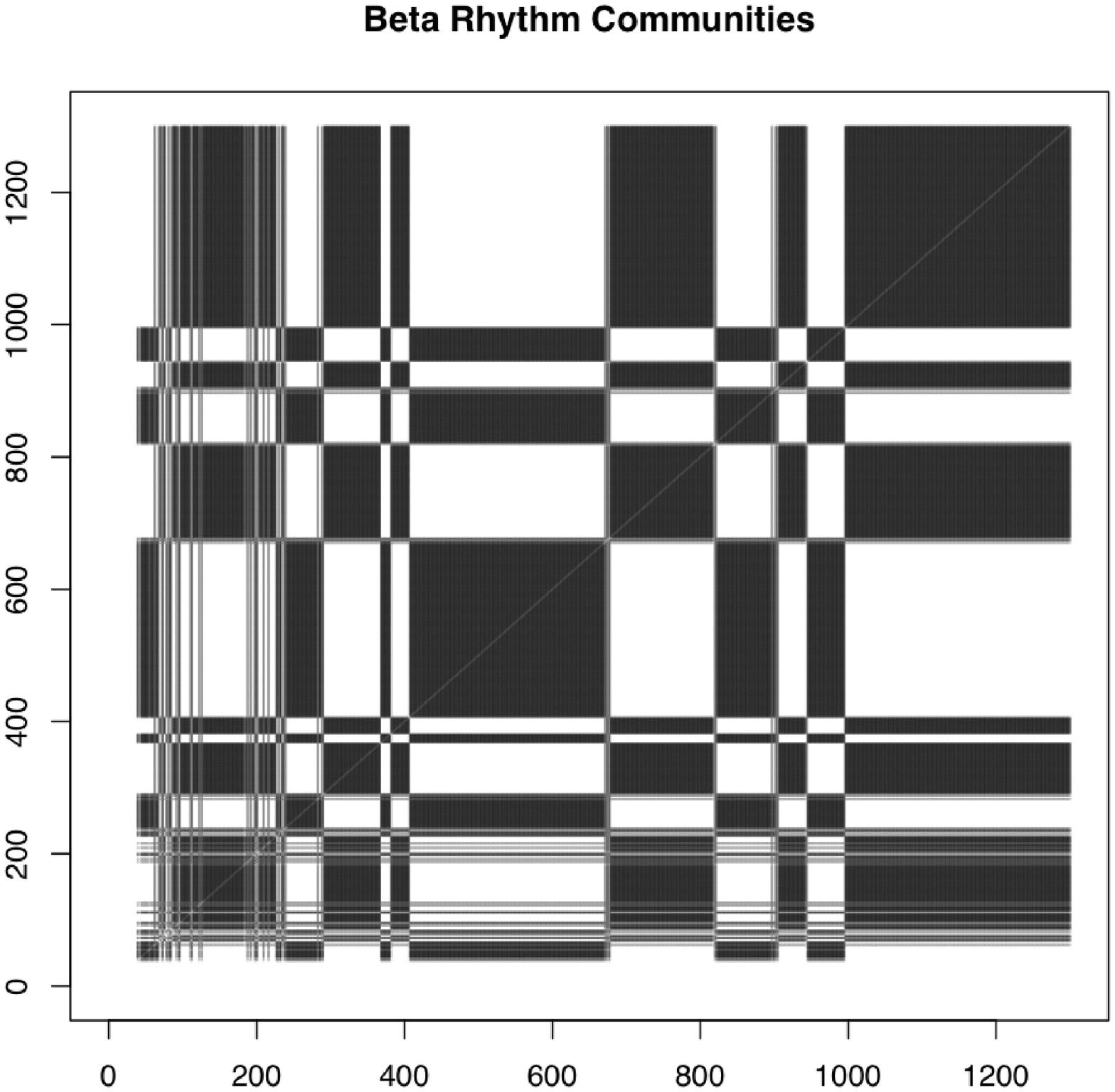}}\end{center}
\caption{Division of the network into functional communities.  Black points
  denote pairs of nodes which are both members of a given community.  During
  the gamma rhythm ($a$) the interneurons (numbered 1001 to 1300) form the core
  of a single community, along with some of the active pyramidal neurons;
  because of the spatially modulated input received by the latter, however,
  some of them belong to another community.  During beta rhythm ($b$), the
  communities re-organize, and in particular formerly inactive pyramidal
  neurons are recruited into the community centered on the interneurons, as
  suggested by the rastergrams.}
\label{fig:communities}
\end{figure}

These expectations are abundantly fulfilled (Fig.\ \ref{fig:communities}).  We
identified communities by running the Reichardt-Bornholdt algorithm with the
maximum number of communities (spin states) set to 25, the modularity
Hamiltonian, and $\gamma = 1$.  (Results were basically unchanged at 40 or 100
spin values.)  In both regimes, there are two overwhelmingly large communities,
containing almost all of the neurons which actually fired, and a handful of
single-neuron communities.  The significant change, visible in the figure, is
in the organization of these communities.

During the gamma rhythm, the 300 interneurons form the core of the larger of
these two communities, which also contains 199 pyramidal neurons.  Another 430
pyramidal neurons belong to a second community.  A final 5 pyramidal cells are
in single-neuron communities; the rest do not fire at all.  A hierarchical
analysis (not shown) has the two large communities merging into a single
super-community.  The regular alternation of the two communities among the
pyramidal neurons, evident in Fig.\ \ref{fig:communities}$a$, is due to the
fact that the external current driving the pyramidal neurons is not spatially
uniform.

With the switch to the beta rhythm, the communities grow and re-organize.  The
community centered on the interneurons expands, to 733 neurons, largely by
incorporating many low-index pyramidal neurons which had formerly been silent,
and are now somewhat erratically synchronized, into its periphery.
Interestingly, many of the latter are only weakly coherent with any one
interneuron (as can be seen by comparing Figs.\ \ref{fig:heatmap}$b$ and
\ref{fig:communities}$b$).  What is decisive is rather their stronger over-all
pattern of coordination with the interneurons, shown by sharing a common
(approximate) firing period, which is half that of the high-index pyramidal
cells (Fig.\ \ref{fig:rastergram}$b$).  Similarly, the other large community,
consisting exclusively of pyramidal neurons, also grows (to 518 members), again
by expanding into the low-index part of the network; there is also considerable
exchange of high-index pyramidal cells between the two communities.  Finally,
nine low-index neurons, which fire only sporadically, belong in clusters of one
or two cells.

\section{Discussion and Conclusion}
\label{sec:discussion}

\subsection{Limitations and Related Approaches}
\label{sec:limitations-and-relations}

Our method is distinguished from earlier work on functional connectivity
primarily by our strong notion of functional community or module, and
secondarily by our measure of functional connectivity.  Previous approaches to
functional connectivity (reviewed in
\cite{Sporns-Tononi-Edelman-theoretical-neuroanatomy,Friston-beyond-phrenology})
either have no concept of functional cluster, or use simple agglomerative
clustering \cite{Hand-Mannila-Smyth}; their clusters are just groups of nodes
with pairwise-similar behavior.  We avoid agglomerative clustering for the same
reason it is no longer used to find anatomical communities: it is insensitive
to the global pattern of connectivity, and fails to divide the network into
coherent components.  Recall (Sect.\ \ref{sec:method-summary}) that every
functional community we find has more intra-cluster information sharing than is
expected by chance, and less inter-cluster information sharing.  This is a
plausible formalization of the intuitive notion of ``module'', but
agglomeration will not, generally, deliver it.

As for using informational coherence to measure functional connectivity, we
discussed its advantages over other measures in Sect.\ \ref{sec:info-coh}
above, and at more length in \cite{Info-Coh-for-NIPS}.  Previous work on
functional connectivity has mostly used surface features to gauge connectivity,
such as mutual information between observables.  (Some of the literature on
clustering general time series, e.g.
\cite{Smyth-clustering-using-HMMs,Oates-Firoiu-Cohen-clustering-with-HMMs,%
  Cadez-Gaffney-Smyth-clustering-individuals}, uses hidden Markov models to
extract latent features, but in a mixture-model framework very different from
our approach.)  The strength of informational coherence is that it is a
domain-neutral measure of nonlinear, stochastic coordination; its weakness is
that it requires us to know the temporal sequence of predictive states of all
nodes in the network.

This need to know the predictive states of each node is the major limitation of
our method.  For some mathematical models, these states are analytically
calculable, but in most cases they must be learned from discrete-value,
discrete-time (``symbolic'') time series.  Those series must be fairly long;
exactly how long is an on-going topic of investigation\footnote{\CSSR\
  converges on the true predictive states (see the appendix), but the rate of
  convergence is not yet known.}, but, empirically, good results are rare with
less than a few thousand time steps.  Similarly, reliable estimates of the
mutual information and informational coherence also require long time series.

Predictive states can be mathematically defined for continuous-value,
continuous-time systems \cite{Knight-predictive-view}, but all current
algorithms for discovering them, not just \CSSR, require symbolic time series.
(Devising a state-reconstruction procedure for continuous systems is another
topic of ongoing research.)  Spike trains, like e-mail networks
\cite{Eckmann-Moses-Sergi-dialog}, are naturally discrete, so this is not an
issue for them, but in most other cases we need to find a good symbolic
partition first, which is non-trivial \cite{Hirata-et-al-estimating-partition}.
The need for long symbolic time series may be especially difficult to meet with
social networks.

\subsection{Directions for Future Work}
\label{sec:future-work}

Our results on the model of hippocampal rhythms, described in the previous
section, are quite promising: our algorithm discovers functional communities
whose organization and properties make sense, given the underlying
micro-dynamics of the model.  This suggests that it is worthwhile to apply the
method to systems where we lack good background knowledge of the functional
modules.  Without pre-judging the results of those investigations, however, we
would like to highlight some issues for future work.

1. Our method needs the full matrix of informational coherences, which is an
$O(N^2)$ computation for a network of size $N$.  If we are interested in the
organization of only part of the network, can we avoid this by defining a local
community structure, as was done for anatomical connectivity by
\cite{Clauset-local-community}?  Alternatively, if we know the anatomical
connectivity, can we restrict ourselves to calculating the informational
coherence between nodes which are anatomically tied?  Doing so with our model
system led to basically the same results (not shown), which is promising; but
in many real-world systems the anatomical network is itself uncertain.

2. The modularity Hamiltonian of Sect.\ \ref{sec:community-discovery} measures
how much information each node shares with other members of its community on a
pairwise basis.  However, some of this information could be redundant across
pairs.  It might be better, then, to replace the sum over pairs with a
higher-order coherence.  The necessary higher-order mutual informations are
easily defined \cite{Sporns-Tononi-Edelman-theoretical-neuroanatomy,%
  Amari-hierarchical-info-geo,Schneidman-et-al-network-information}, but the
number of measurements needed to estimate them from data grows exponentially
with the number of nodes.  However, it may be possible to approximate them
using the same Chow-Liu bounds employed by \cite{Info-Coh-for-NIPS} to estimate
the global coherence.

3. It would be good if our algorithm did not simply report a community
structure, but also assessed the likelihood of the same degree of modularity
arising through chance, i.e., a significance level.  For anatomical
communities, Guimera et al.\ \cite{Guimera-et-al-modularity-from-fluctuations}
exploit the spin-system analogy to show that random graph processes without
community structure will nonetheless often produce networks with non-zero
modularity, and (in effect) calculate the sampling distribution of Newman's $Q$
using both Erd\"os-R{\'e}nyi and scale-free networks as null models.  (See
however \cite{Reichardt-Bornholdt-truly-modular} for corrections to their
calculations.)  To do something like this with our algorithm, we would need a
null model of functional communities.  The natural null model of functional
connectivity is simply for the dynamics at all nodes to be independent, and
(because the states are Markovian) it is easy to simulate from this null model
and then bootstrap $p$-values.  We do not yet, however, have a class of
dynamical models where there nodes share information, but do so in a completely
distributed, a-modular way.

4. A variant of the predictive-state analysis that underlies informational
coherence is able to identify coherent structures produced by spatiotemporal
dynamics \cite{Automatic-Filters}.  Moreover, these techniques can be adapted
to network dynamics, if the anatomical connections are known.  This raises
numerous questions.  Are functional communities also coherent structures?  Are
coherent structures in networks \cite{Moreira-et-al-system-wide-coordination}
necessarily functional communities?  Can the higher-order interactions of
coherent structures in regular spatial systems be ported to networks, and, if
so, could functional re-organization be described as a dynamical process at
this level?

\subsection{Conclusion}

Network dynamical systems have both anatomical connections, due to persistent
physical couplings, and functional ones, due to coordinated behavior.  These
are related, but logically distinct.  There are now many methods for using a
network's anatomical connectivity to decompose it into highly modular
communities, and some understanding of these methods' statistical and
statistical-mechanical properties.  The parallel problem, of using the pattern
of functional connectivity to find functional communities, has scarcely been
explored.  It is in many ways a harder problem, because measuring functional
connectivity is harder, and because the community organization is itself
variable, and this variation is often more interesting than the value at any
one time.

In this paper, we have introduced a method of discovering functional modules in
stochastic dynamical networks.  We use informational coherence to measure
functional connectivity, and combine this with a modification of the
Potts-model community-detection procedure.  Our method gives good results on a
biophysical model of hippocampal rhythms.  It divides the network into two
functional communities, one of them based on the inhibitory interneurons, the
other consisting exclusively of excitatory pyramidal cells.  The two
communities change in relative size and re-organize during the switch from
gamma to beta rhythm, in ways which make sense in light of the underlying model
dynamics.  While there are theoretical issues to explore, our success on a
non-trivial simulated network leads us to hope that we have found a general
method for discovering functional communities in dynamic networks.

\section*{Acknowledgments}

Thanks to L. A. N. Amaral, S. Bornholdt, A. Clauset, R. Haslinger, C. Moore and
M. E. J. Newman for discussing community discovery and/or network dynamics; to
J. Reichardt for valuable assistance with the implementation of his algorithm
and general suggestions; and to the editors for their patience.

\appendix

\section{The \CSSR\  Algorithm}
\label{app:CSSR}

This appendix briefly describes the \CSSR\ algorithm we use to reconstruct the
effective internal states of each node in the network.  For details, see
\cite{CSSR-for-UAI}; for an open-source C++ implementation, see
\texttt{http://bactra.org/CSSR/}.  For recent applications of the algorithm to
problems in crystallography, anomaly detection and natural language processing,
see \cite{Varn-infinite-range-order,Ray-anomaly-detection,%
  Padro-Padro-FSMNLP05, Padro-Padro-PLN05,Padro-Padro-RANLP05}.

We wish to predict a dynamical system or stochastic process
$\left\{X_t\right\}$.  By $X^t_s$ we will denote the whole trajectory of the
process from time $s$ to time $t$, inclusive, by $X_t^+$ the whole ``past'' or
``history'' of the process through time $t$, and by $X_t^+$ its ``future'', its
trajectory at times strictly greater than $t$.  The ``state'' of
$\left\{X_t\right\}$ at time $t$ is a variable, $S_t$, which fixes the
distribution of all present or future observations, i.e., the distribution of
$X^{+}(t)$ \cite{CMPPSS,predictive-representations-of-state}.  As such, the
state is a minimal sufficient statistic for predicting the future of the
process.  Sufficiency is equivalent to the requirement that $I[X^{+}_t;X^{-}_t]
= I[X^{+}_t;S_t]$, where $I[\cdot;\cdot]$ is the mutual information
\cite{Kullback-info-theory-and-stats}.  In general, $S_t = \epsilon(X^{-}_t)$,
for some measurable functional $\epsilon(\cdot)$ of the whole past history of
the process up to and including time $t$.  If $\left\{X_t\right\}$ is
Markovian, then $\epsilon$ is a function only of $X_t$, but in general the
state will incorporate some history or memory effects.  Each state, i.e.,
possible value of $\epsilon$, corresponds to a predictive distribution over
future events, and equally to an equivalence class of histories, all of which
lead to that conditional distribution over future events.  State-reconstruction
algorithms use sample paths of the process to find approximations
$\hat{\epsilon}$ to the true minimal sufficient statistic $\epsilon$, and
ideally the approximations converge, at least in probability.  The \CSSR\ 
algorithm \cite{CSSR-for-UAI} does so, for discrete-valued, discrete-time,
conditionally-stationary processes.

\CSSR\  is based on the following result about predictive sufficiency
\cite[pp.\ 842--843]{CMPPSS}.  Suppose that $\epsilon$ is next-step sufficient,
i.e., $I[X_{t+1};X^{-}_t] = I[X_{t+1};\epsilon(X^{-}_t)]$, and that it can be
updated recursively: for some measurable function $T$, $\epsilon(X^{-}_{t+1}) =
T(\epsilon(X^{-}_t,X_{t+1})$.  Then $\epsilon$ is predictively sufficient for
the whole future of the process --- intuitively, the recursive updating lets us
chain together accurate next-step predictions to go as far into the future as
we like.  \CSSR\  approximates $\epsilon$ by treating it as a partition, or set
of equivalence classes, over histories, and finding the coarsest partition
which meets both of the conditions of this theorem.  Computationally, \CSSR\ 
represents states as sets of suffixes, so a history belongs to a state
(equivalence class) if it terminates in one of the suffixes in that state's
representation.  That is, a history, $x^-_t$, will belong to the class $C$,
$x^{-}_t \in C$, if $x^t_{t-|c|+1} = c$, for some suffix $c$ assigned to $C$,
where $|c|$ is the length of the suffix.\footnote{The algorithm ensures that
  there are never overlapping suffixes in distinct states.}

In the first stage, \CSSR\ tries to find a partition of histories which is
sufficient for next-step prediction.  It begins with the trivial partition, in
which all histories belong to the same equivalence class, defined by the null
suffix (corresponding to an IID process), and then successively tests whether
longer and longer suffices give rise to the same conditional distribution for
the next observation which differ significantly from the class to which they
currently belong.  That is, for each class $C$, suffix $c$ in that class, and
possible observable value $a$, it tests whether $\Prob{X_{t+1}|X^-_t \in C}$
differs from $\Prob{X_{t+1}|X^t_{t-|c|+1} = c, X_{t-|c|} = a}$.  (We use
standard tests for discrepancy between sampled distributions.)  If an extended,
child suffix ($ac$) does not match its current classes, the parent suffix ($c$)
is deleted from its class ($C$), and \CSSR\ checks whether the child matches
any existing class; if so it is re-assigned to the closest one, and the
partition is modified accordingly.  Only if a suffix's conditional distribution
($\Prob{X_{t+1}|X^t_{t-l} = ac}$) differs significantly from all existing
classes does it get its own new cell in the partition.

The result of this stage is a partition of histories (i.e., a statistic) which
is close to being next-step sufficient, the sense of ``close'' depending on the
significance test.  In the second stage, \CSSR\  iteratively refines this
partition until it can be recursively updated.  This can always be done, though
it is potentially the most time-consuming part of the algorithm\footnote{In
  terms of automata theory, recursive updating corresponds to being a
  ``deterministic'' automaton, and non-deterministic automata always have
  deterministic equivalents.}.  The output of \CSSR, then, is a set of states
which make good next-step predictions and can be updated recursively, and a
statistic $\hat{\epsilon}$ mapping histories to these states.

If the true number of predictive states is finite, and some mild technical
assumptions hold \cite{CSSR-for-UAI}, a large deviations argument shows that
$\Prob{\hat{\epsilon} \neq \epsilon} \rightarrow 0$ as the sample size $n
\rightarrow \infty$.  That is, \CSSR\  will converge on the minimal sufficient
statistic for the data-generating process, even though it lacks an explicit
minimization step.  Furthermore, once the right statistic has been discovered,
the expected $L_1$ (total variation) distance between the actual predictive
distribution, $\Prob{X^{+}_t|\epsilon(X^{-}_t)}$ and that forecast by the
reconstructed states, $\Prob{X^{-}_t|\hat{\epsilon}(X^{-}_t)}$, goes to zero
with rate $O(n^{-1/2})$, which is the same rate as for IID data.  The time
complexity of the algorithm is at worst $O(n) + O(k^{2L+1})$, where $k$ is the
number of discrete values possible for $X_t$, and $L$ is the maximum length of
suffices considered in the reconstruction.  Empirically, average-case time
complexity is much better than this.

\bibliographystyle{splncs} \bibliography{locusts}

\end{document}